\documentclass[9pt,twocolumn,twoside]{osajnl}

\usepackage{placeins}

\journal{ol} 

\setboolean{shortarticle}{true} 
\usepackage{siunitx}

\ifthenelse{\boolean{shortarticle}}{\colorlet{color2}{color2b}}{\colorlet{Colorado}{color2}}

\title{Design study of random spectrometers for applications at optical frequencies}

\author[1*]{Paris Varytis}
\author[2]{Dan{-}Nha Huynh}
\author[3,4]{Wladislaw Hartmann}
\author[3,4]{Wolfram Pernice}
\author[1,2]{Kurt Busch}

\affil[1]{Max-Born-Institut, Max-Born-Str. 2A, 12489 Berlin, Germany}
\affil[2]{Humboldt-Universit\"at zu Berlin, Institut f\"ur Physik, 
          AG Theoretische Optik \& Photonik, Newtonstr 15, 12489 Berlin, Germany}
\affil[3]{University of M\"unster, Institute of Physics, Wilhelm-Klemm-Str. 10,
          48149 M\"unster, Germany}
\affil[4]{University of M\"unster, CeNTech - Center for Nanotechnology,
          Heisenbergstr. 11, 48149 M\"unster, Germany}

\affil[*]{Corresponding author: varytis@physik.hu-berlin.de}

\ociscodes{(290.1990) Diffusion; 
           (290.4020) Mie theory; 
           (290.4210) Multiple scattering; 
           (300.6190) Spectrometers}

\begin{abstract}
Compact spectrometers based on disordered planar waveguides exhibit a rather high
resolution with a relatively small footprint as compared to conventional spectrometers.
This is achieved by multiple scattering of light which -- if properly engineered --
significantly enhances the effective optical path length. Here a design study of random
spectrometers for TE- and TM-polarized light is presented that combines the results
of Mie
theory, multiple-scattering theory and full electromagnetic simulations. It is shown
that
the performance of such random spectrometers depends on single scattering quantities,
notably on the overall scattering efficiency and the asymmetry parameter. Further, the
study shows that a well-developed diffusive regime is not required in practice and
that a
standard integrated-optical layout is sufficient to obtain efficient devices even for
rather weakly scattering systems consisting of low index inclusions in high-index
matrices
such as pores in planar silicon-nitride based waveguides. This allows for both
significant reductions in footprint with acceptable losses in resolution and for device
operation in the visible and near-infrared frequency range.
\end{abstract}

\setboolean{displaycopyright}{false}

\begin{document}

\maketitle

\thispagestyle{empty}
\pagestyle{empty}

Random spectrometers are suitable for portable sensing and efficient lab-on-a-chip
functionality. While previous efforts aimed at the telecom wavelength regime with
silicon-based structures~\cite{redding2013compact}, 
random spectrometers operating in the near-infrared (NIR) spectral region where water, 
cells and tissue exhibit low absorption (and molecules feature low fluorescence) could 
offer important advantages for biological and medical applications such as Raman 
spectroscopy~\cite{dhakal2016single} 
and bioimaging~\cite{yodh1995spectroscopy, gayen1996emerging}. 
Similar arguments can be made with respect to operations at visible frequencies.

In the present work, we analyze the performance of planar random spectrometers for
frequencies in the visible, NIR and telecom ranges for both polarizations (TE and
TM) by
means of Mie and multiple-scattering theory in conjunction with the numerical
solution of
the Maxwell equations. To provide the required broadband optical transparency, we
conduct
our design study for silicon-nitride based waveguides.
In fact, silicon nitride is a robust material~\cite{zhao2015visible, moss2013new}
with low thermal
sensitivity that is compatible with CMOS processes for low-cost mass fabrication.
Further,
silicon nitride is transparent in the wavelength regime between \SI{700} {\nano \meter}
and \SI{1100} {\nano \meter} where it features a relatively high refractive index
($n\sim2$) so that it is suitable for many applications in the visible and NIR
regime~\cite{goykhman2010ultrathin,subramanian2012near,
romero2013silicon,romero2013visible}. As scattering area, we consider a set
of identical pores that have been etched into the silicon nitride waveguide. 

\begin{figure}[htbp]
\centerline{\includegraphics*[width=\linewidth]{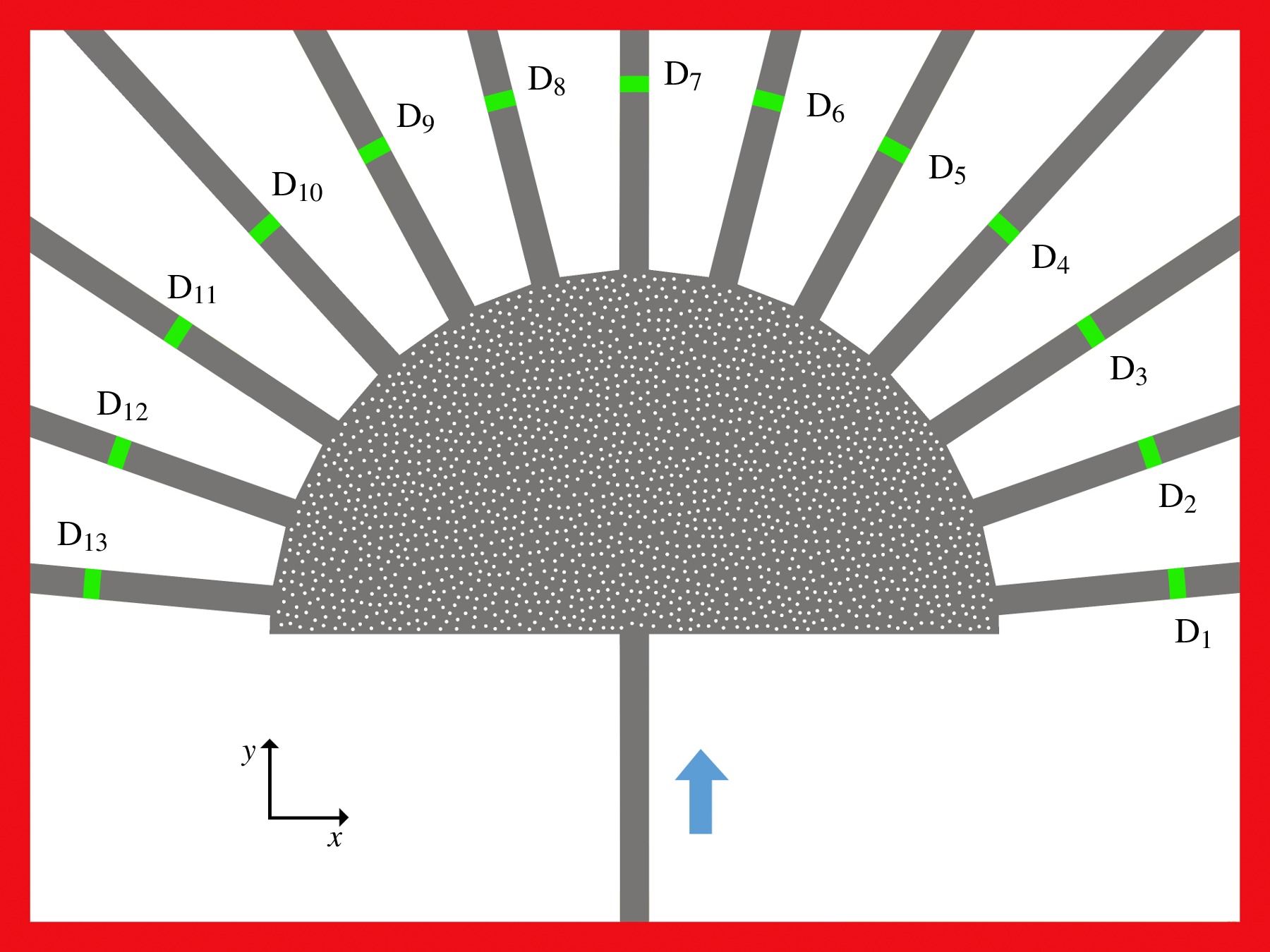}} 
\caption{ (Color online) Schematic of the random spectrometer layout considered
  in this work. Radiation is launched into the center of the semicircular
  scattering area through an input waveguide and after multiple scattering 
  reaches the 13 output waveguides where it is detected at the detectors
  D$_1$, ... D$_{13}$ (green lines), or leaks into free space. Grey-shaded regions 
  correspond to silicon nitride, white-shaded regions correspond to free 
  space, and red-shaded regions represent perfectly matched layers that 
absorb any outgoing radiation. See text for further details.}
\label{fig1}
\end{figure}

In Fig.~\ref{fig1} we depict the considered random spectrometer geometry: We use a
semicircular scattering area with a radius of $L=\SI{25} {\micro \meter}$, which
consists
of a random array of identical pores with \SI{125} {\nano \meter} radius that cover
9 \%
of the scattering area. A single input waveguide (with a width of \SI{2}{\micro\meter})
launches light into the center of the scattering area and after multiple scattering
light
reaches the 13 readout waveguides (with widths of \SI{2}{\micro\meter}), which are
placed
with equal angular distance around the outer circumference. The backscattering into the
launch waveguide provides the 14th readout port.
Into this two-dimensional system, we launch a fundamental mode pulse, and at the
locations of
the detectors D$_1$, ..., D$_{13}$ we determine the radiation that exits through the
readout port.
After a Fourier transform of the (time-dependent) output signals,
we recover the frequency-resolved transmission matrix $T$ with a single time-domain
computation. For the actual numerical computations we utilize a Discontinuous Galerkin
Time-Domain (DGTD) finite-element approach
~\cite{busch2011discontinuous,niegemann2012efficient}.

To quantify the spectral resolution of the spectrometer we compute the spectral
correlation function by
~\cite{redding2013compact}
\begin{equation}
  \label{SP}
  C(\Delta \lambda,D_i) 
  = 
  \dfrac{\langle I(\lambda,D_i) I(\lambda+ \Delta \lambda,D_i) 
  \rangle}{\langle I(\lambda,D_i) \rangle 
  \langle I(\lambda+ \Delta \lambda,D_i) \rangle} 
  - 1~,
\end{equation}
where $I( \lambda , D_i)$ denotes the light intensity at wavelength $\lambda$ and
detector
$D_i$ ($i = 1,2, ...,13$) and the average is taken over the wavelength. $C$ is
averaged over all detectors and
normalized at $\Delta \lambda=0$. As a result, the half-width at half-maximum (HWHM) of
the correlation function provides an estimate of the spectral resolution.
\FloatBarrier

Since the pore filling fraction is $f=0.09$, multiple scattering can be treated 
via the independent scattering approximation
~\cite{van1991speed, van1992speed}, 
where the multiple scattering process can be described by approriate combinations
of single-scattering quantities (we will touch upon the case of higher filling
ratios below).
The scattering efficiency for an infinite homogeneous cylindrical scatterer at
normal incidence reads
\begin{equation}
  \label{Q}
  Q^\mathrm{s,p}_\mathrm{sc} 
  =
  \dfrac{2}{kR}\Big[\vert a_{0}^\mathrm{s,p}\vert^{2} 
  +
  2\sum_{n=1}^{\infty}\vert a_{n}^\mathrm{s,p}\vert^{2}\Big]~,
\end{equation}
where $k$ is the wavenumber in the background material, $R$ denotes the radius 
of the cylinder, $a_{n}$ ($n = 0,1,2, ...$) represent the Mie scattering coefficients. 
Finally, the superscripts $\mathrm s$ and $\mathrm p$ correspond, respectively, to
$\mathrm s$-wave 
(TM-polarization, electric field perpendicular to the $xy$-plane in
Fig.~\ref{fig1}) 
and $\mathrm p$-wave (TE-polarization, electric field parallel to the $xy$-plane
in Fig.~\ref{fig1}). 
The associated differential scaterring efficiency is
\begin{equation}
   \label{SA}
   \dfrac{\partial Q^\mathrm{s,p}_\mathrm{sc}(\phi)}{\partial \phi}
   =
   \dfrac{2}{\pi kR} \bigg\vert a_{0}^\mathrm{s,p} 
   +
   2\sum_{n=1}^{\infty} a_{n}^\mathrm{s,p}cos(n \theta)\bigg\vert^{2}~,
\end{equation}
where $\theta=\pi -\phi$ is the scattering angle. 
The average cosine of the scattering angle defines
~\cite{hulst1957light,bohren2008absorption} 
the so-called asymmetry parameter $g$:
\begin{equation}
  \label{g}
  \begin{aligned}
    g^\mathrm{s,p}=\langle \cos(\theta) \rangle
    &=
    \dfrac{ \int_{0}^{\pi} 
      \dfrac{\partial Q^\mathrm{s,p}_\mathrm{sc}(\phi)}{\partial \phi} 
    \cos(\theta) d\theta} { Q^\mathrm{s,p}_\mathrm{sc}} \\ 
    &=
    \dfrac{4}{kRQ^\mathrm{s,p}_\mathrm{sc}}\sum_{n=0}^{\infty}
    a^\mathrm{s,p}_{n}(a^\mathrm{s,p}_{n+1})^{*}~.
  \end{aligned}
\end{equation}
For isotropic scattering, the asymmetry parameter vanishes and for predominantly 
forward (backward) scattering the asymmetry parameter takes on positive (negative)
values.
From multiple-scattering theory 
~\cite{ishimaru1978wave, arruda2016electromagnetic}
it follows that for lossless scatterers the transport mean free path is given by
\begin{equation}
  \label{lt}
  l^\mathrm{s,p}_\mathrm{t} = \dfrac{\pi R}{2fQ^\mathrm{s,p}_\mathrm{sc}(1-g^\mathrm{s,p})}~
\end{equation}
within the independent scattering approximation~\cite{van1991speed, van1992speed}.
In the diffusive regime, the spectral resolution of random spectrometers scales 
as $l^\mathrm{s,p}_\mathrm{t} / L^{2}$
~\cite{pine1988diffusing}, 
so that reduced transport mean free paths $l^\mathrm{s,p}_\mathrm{t}$ are generally
desirable. 
According to Eq.~\ref{lt}, short transport mean free paths can be expected when 
the (wavelength-dependent) scattering efficiency is high (strong scattering) and 
when the asymmetry parameter is significantly below 1 (near isotropic or even
predominant backward scattering). 
\begin{figure}[htbp]
\centerline{\includegraphics*[width=\linewidth]{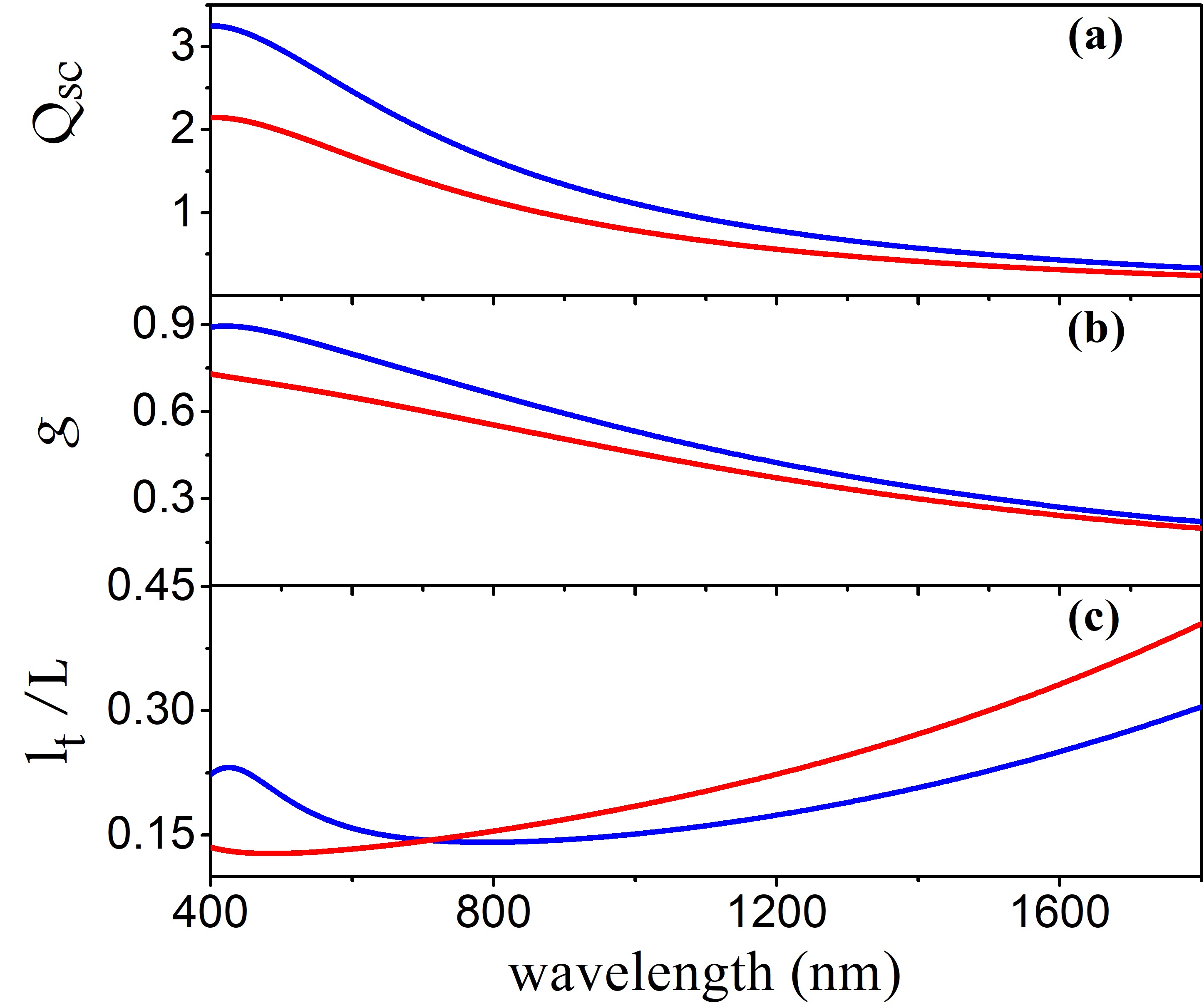}} 
\caption{ (Color online) (a) Scattering efficiency $Q_\mathrm{sc}$ and (b) asymmetry
parameter 
          $g$ of an air cylinder of radius 125~nm embedded in a silicon nitride
matrix. 
          The system is irradiated with $\mathrm s$- and $\mathrm p$-waves under normal incidence
          (blue and red lines, respectively). 
          (c) Calculated transport mean free path $l_\mathrm{t}$ normalized to the radius $L$ of
                                        the scattering area of a 2D disordered medium composed of air pores in a 
                                        silicon nitride matrix where the pores occupy a fraction $f=0.09$ of the 
                                        available area.}
\label{fig2}
\end{figure}
\FloatBarrier
For an air pore of \SI{125} {\nano \meter} radius 
embedded in a silicon-nitride matrix, we display in Fig.~\ref{fig2} both, the
scattering efficiency and the asymmetry parameter, for TE and TM polarization.
We observe that when moving from telecom frequencies to visible frequencies, the 
scattering efficiencies increase monotonously while simultaneously the asymmetry 
parameter concurrently moves to even stronger forward scattering. Despite these
opposing trends, the overall effect is that the transport mean free path reduces
when moving from telecom to visible wavelengths (with a minimum around \SI{700} 
{\nano \meter} for TE polarization) so that high spectral resolution of the
random spectrometer is expected for the red end of the visible spectrum and NIR 
frequencies. Further reduction of the operation wavelength ($\sim \SI{400} {\nano
\meter}$) would lead to an elongated transport mean free path for TE polarization 
due to a stronger increase in the forward scattering characteristics relative to
the corresponding increase in the scattering efficiency.

\begin{figure}[t!]
\centerline{\includegraphics*[width=\linewidth]{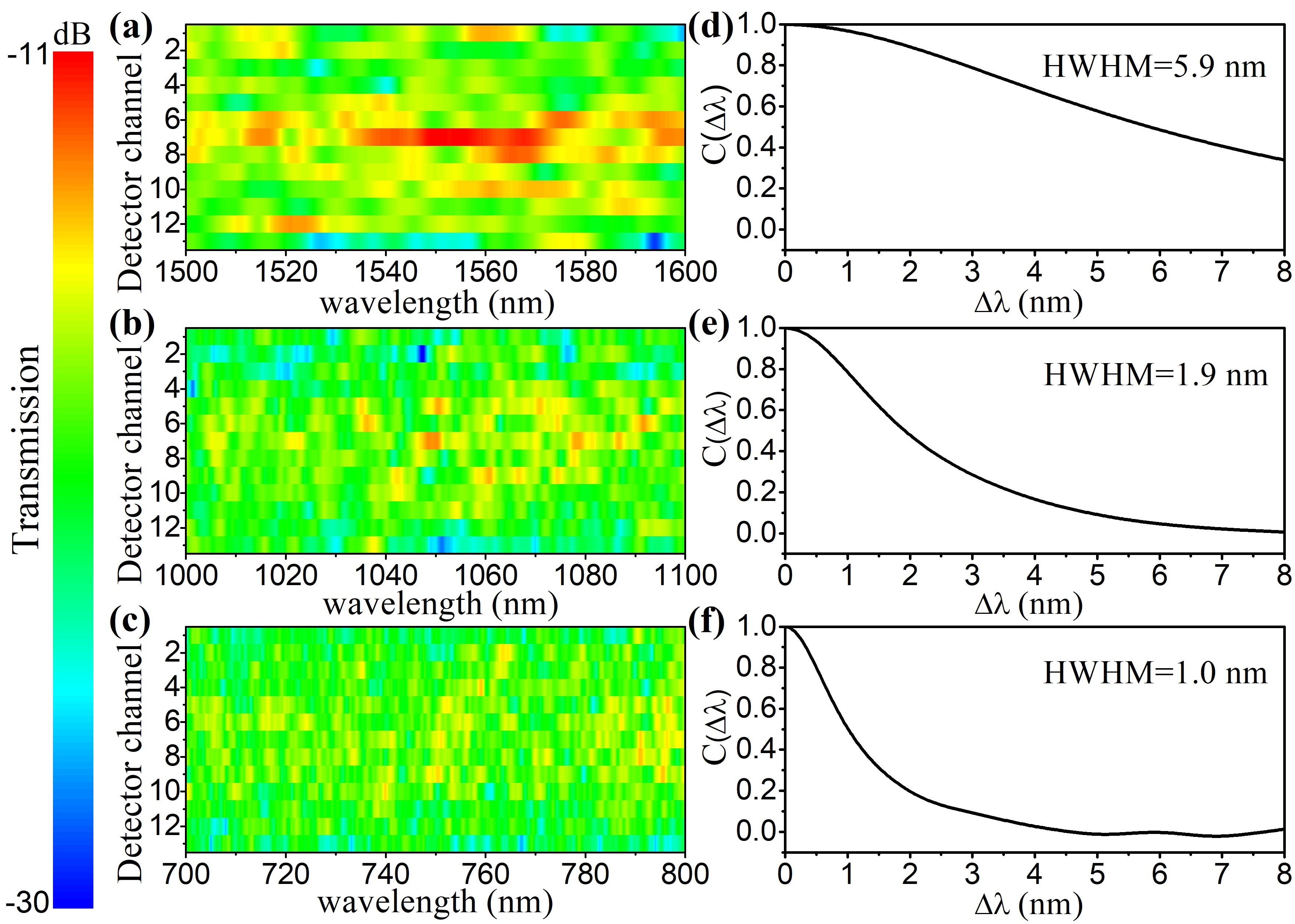}} 
\caption{(Color online) Panels (a), (b) and (c): Wavelength- and detector-resolved
transmission 
         matrix for the random spectrometer depicted in Fig.~\ref{fig1} for TE
polarization 
                                 for the visible, NIR and telecom wavelength region, respectively. Light is
launched 
                                 from the input port and is detected by the detectors D$_{1}$ to D$_{13}$. The ordinate
labels
                                 correspond to the detector index. The color coding corresponds to the relative
                                 intensity transmitted into the different waveguides and expressed in decibels.
                                 Panels (d), (e) and (f): The normalized spectral correlation function
corresponding to the wavelength
                                 regimes of panels (a), (b) and (c).}
\label{fig3}
\end{figure}

\begin{figure}[t!]
\centerline{\includegraphics*[width=\linewidth]{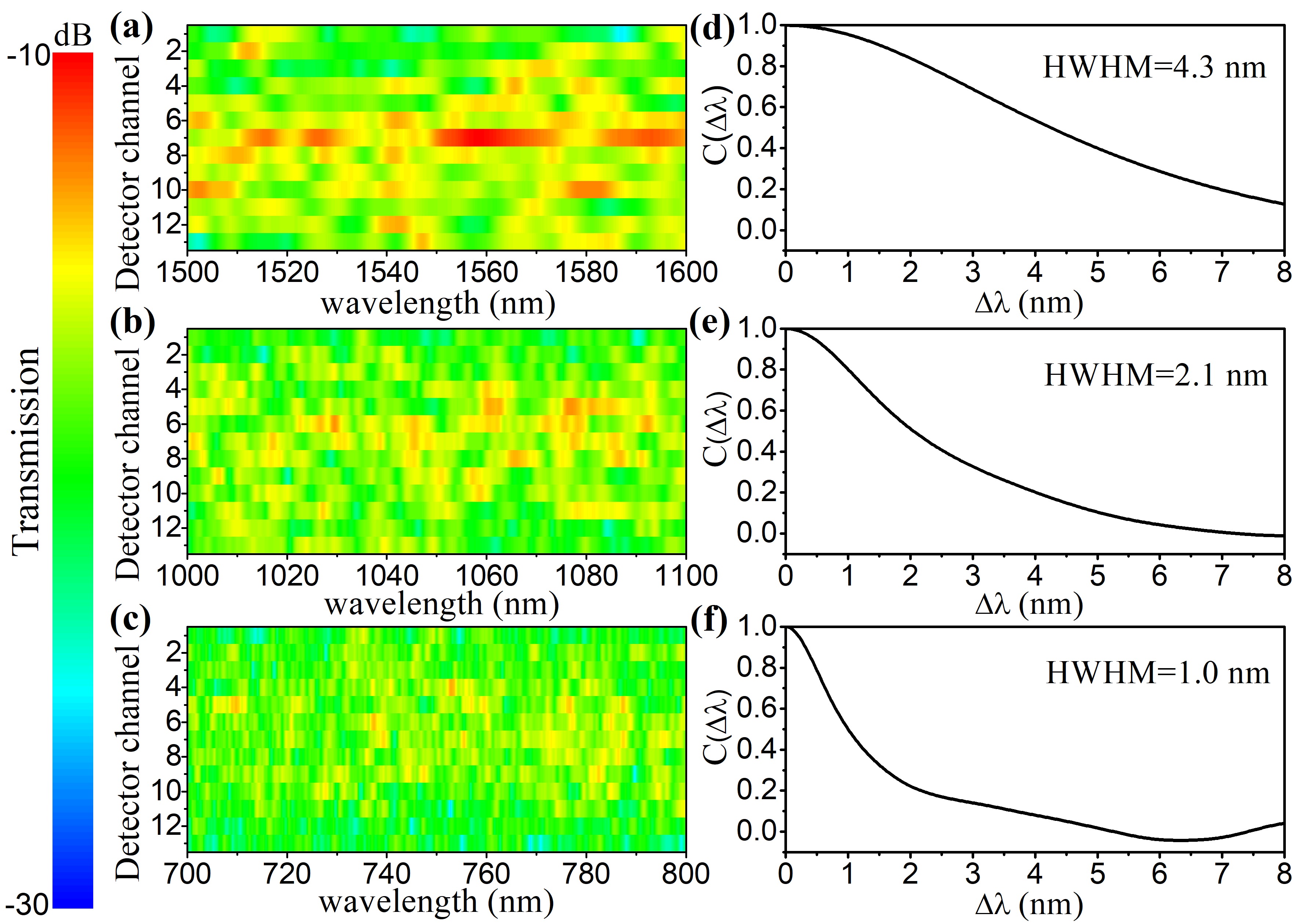}} 
\caption{(Color online) Panels (a), (b) and (c): Wavelength- and detector-resolved
transmission 
         matrix for the random spectrometer depicted in Fig.~\ref{fig1} for TM
polarization 
                                 for the visible, NIR and telecom wavelength region, respectively. Light is
launched 
                                 from the input port and is detected by the detectors D$_{1}$ to D$_{13}$. The ordinate
labels
                                 correspond to the detector index. The color coding corresponds to the relative
                                 intensity transmitted into the different waveguides and expressed in decibels. 
                                 Panels (d), (e) and (f): The normalized spectral correlation function
corresponding to the wavelength
                                 regimes of panels (a), (b) and (c).}
\label{fig4}
\end{figure}

In Figs.~\ref{fig3} and~\ref{fig4}, we display the wavelength- and detector-resolved
transmission matrices for TE and TM polarization, respectively.  As described above,
these
results have been obtained for the random spectrometer sketched in Fig.~\ref{fig1} by
solving the Maxwell equations numerically using the DGTD upwind scheme
approach~\cite{busch2011discontinuous}. Within this approach we use an adaptive
tetrahedron mesh for the spatial discretization with a minimal element size of
\SI{4.7} {\nano \meter} and a
polynomial order of 3. Meanwhile, the time-stepping is handled by a 14-stage fourth
order
low-storage-Runge scheme~\cite{niegemann2012efficient}.

At telecom wavelengths (\SI{1500} {\nano \meter}-\SI{1600} {\nano \meter}), we find 
that the transport mean free path becomes comparable to the radius of the scattering 
area so that ballistic transport is dominant. As a result, the transmission matrix 
(Fig.~\ref{fig3}(a) for TE polarization and Fig.~\ref{fig4}(a) and (b) for TM
polarization), 
is characterized by concentrations of the intensity in the central output ports (6-8).

\begin{figure}[h]
  \centerline{\includegraphics*[width=\linewidth]{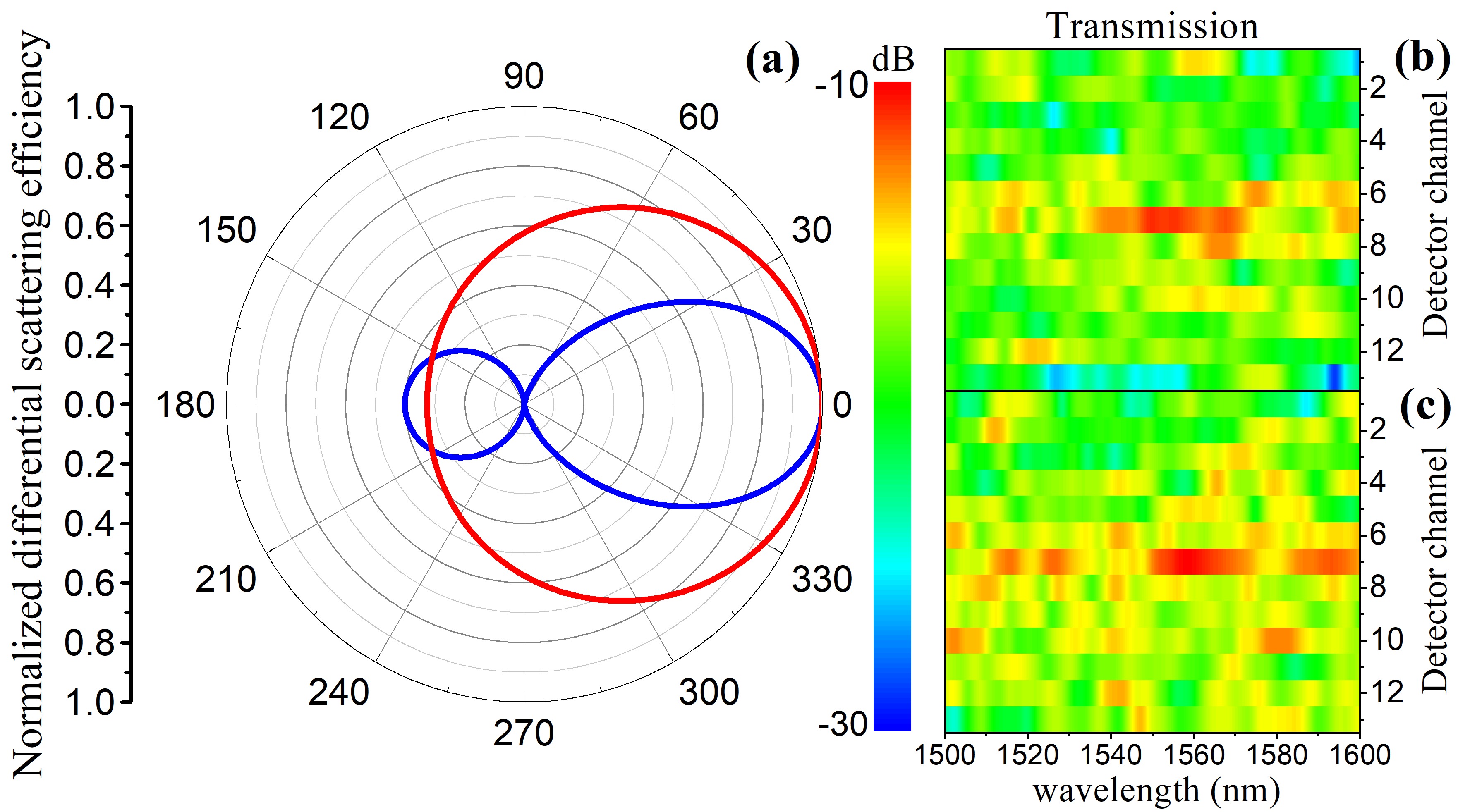}} 
  \caption{(Color online) Panel (a): Angular scattering characteristics,
Eq.~(\ref{SA}),
    for an air pore of 125~nm radius embedded in a silicon nitride matrix, for TE (blue
    line) and TM (red line) polarizations and fixed wavelength $\lambda=1500$~nm. Light
    scatters to larger angles for TM polarization due to predominance of $\mathrm s$-wave
    scattering. Panels (b),(c): Wavelength- and detector-resolved transmission
matrix for TE
  and TM polarizations, respectively.}
  \label{fig5}
\end{figure}

Upon moving to NIR wavelengths (\SI{1000} {\nano \meter}-\SI{1100} {\nano \meter}) 
stronger multiple scattering occurs and we observe the onset of diffusion where the
transmission matrix exhibits a more uniform distribution over the different output 
ports (cf. Fig.~\ref{fig3}(b)), while at the same time the random spectrometer
exhibits 
higher resolutions: Here, the HWHM is \SI{1.9} {\nano \meter} for TE polarization
and  \SI{2.1} {\nano \meter}  for TM polarization, respectively, which is
essentially less than half the values obtained at telecom wavelengths. 
Finally, we would like to note that the differences between TE and TM polarization
in the angular distribution of light over the output ports can explained by the 
angular characteristics of light scattering by a single scatterer (cf.
Fig.~\ref{fig5}). 
For TE polarization, scattering by $\mathrm p$-waves is dominant and light scatters more
into the forward direction so that it is primarily detected in the central output
ports as long as the diffusive regime is not fully developed. For TM polarization,
scattering by $\mathrm s$-waves is dominant, which leads to more isotropic single-particle
scattering and subsequently the light is more evenly distributed over all output
ports even if the diffusive regime is not fully developed (cf. Figs.~\ref{fig3}(b) 
and~\ref{fig4}(b)).

At the red end of the visible spectrum (\SI{700} {\nano \meter}-\SI{800} {\nano
\meter}),
we find that the transport mean free path is further reduced and the transmission 
matrix (Figs.~\ref{fig3}(c) and~\ref{fig4}(c)) is characterized by sharp peaks with
a uniform spatial distribution, i.e., we find diffusive speckle patterns. For this
wavelength range, the random spectrometer exhibits the highest resolution, \SI{1.0} 
{\nano \meter} for both polarizations (see Figs.~\ref{fig3}(f) and~\ref{fig4}(f)).

Clearly, the above numerical results are in full agreement with the predictions of 
the combined Mie and multiple-scattering theory and demonstrate that the highest
spectral resolution of the spectrometer is obtained when the diffusive regime is
fully developed. However, acceptable resolution can already be obtained when the
size of the scattering region just barely allows for the onset of diffusion. If 
ballistic transport is dominant, the spectral resolution will be lower. Thus, which 
transport regime to select, should, thus, be based on a compromise between desired 
resolution and available footprint of the device. 
At this point it should be noted that scattering from low-index inclusions in 
high-index matrices, quite generally, does not exhibit pronounced Mie resonances 
notably at low and moderate index contrasts. This suggests that our findings are 
of a rather general nature and, for instance, apply to other material systems
such as ZnO, chalcogenides and LiNbO$_3$ and broad wavelength ranges. 

In summary, through a combination of electrodynamic simulations, Mie theory and
multiple-scattering theory, we have studied the response of planar random
spectrometers 
in low-footprint integrated-optical layout for TE and TM polarization at telecom, NIR 
and visible wavelengths. We have shown that the attainable spectral resolution depends 
on the interplay of two single scattering properties, scattering efficiency and the
asymmetry parameter. The highest resolution is obtained for systems where the 
diffusive regime is fully developed. However, the transition regime between ballistic
and diffusive transport might be sufficient for certain applications. These results
are based on low filling fractions of the scatterers. For higher filling fractions,
strong multiple-scattering corrections blur the connection to the single-scattering 
quantities. 
In this high-$f$ regime, an effective medium theory capable of reliably determining 
transport mean free paths
~\cite{kirchner1998transport} 
can be employed instead of the independent scattering approximation. Our design 
study provides a basis for the realization of random spectrometers in the visible 
and NIR wavelength range.
\FloatBarrier

\section*{Funding}
We acknowledge support by the Deutsche Forschungsgemeinschaft (DFG) under 
project Bu 1107/10-1 within the framework of the DFG priority program SPP 1839 
({\it Tailored Disorder}).

\bibliography{bibfile}

\end{document}